\documentclass[12pt]{article}

\usepackage{bm}
\usepackage{latexsym}
\usepackage{amsmath, amsthm,amsfonts,url}
\usepackage[]{graphicx}

\expandafter\ifx\csname package@font\endcsname\relax\else
 \expandafter\expandafter
 \expandafter\usepackage
 \expandafter\expandafter
 \expandafter{\csname package@font\endcsname}%
\fi

\newcommand{\ket}[1]{\vert#1\rangle}
\newcommand{\phiplus}{\ket{\phi^{+}}}
\newcommand{\phiminus}{\ket{\phi^{-}}}
\newcommand{\psiplus}{\ket{\psi^{+}}}
\newcommand{\psiminus}{\ket{\psi^{-}}}

\begin{document}

\title{Entanglement swapping, light cones and elements of reality}%

\title{\LARGE{Entanglement swapping, light cones and elements~of~reality}}

\author{
Anne Broadbent \hspace{0.75cm} Andr\'e Allan M\'ethot\\[0.5cm]
\normalsize\sl D\'epartement d'informatique et de recherche op\'erationnelle\\[-0.1cm]
\normalsize\sl Universit\'e de Montr\'eal, C.P.~6128, Succ.\ Centre-Ville\\[-0.1cm]
\normalsize\sl Montr\'eal (QC), H3C 3J7~~\textsc{Canada}\\
{\normalsize\texttt{\{broadbea,\,methotan\}}\textbf{\char"40}\texttt{iro.umontreal.ca}}}

\date{19 July 2006}
\maketitle


\begin{abstract}
Recently, a number of two-participant all-versus-nothing Bell
experiments have been proposed. Here, we give local realistic
explanations  for these experiments. More precisely, we examine the
scenario where a participant swaps his entanglement with two other
participants and then is removed from the experiment; we also
examine the scenario where two particles are in the same light cone,
i.e.\ belong to a single participant. Our conclusion is that, in
both cases,
the proposed experiments are not convincing proofs against local
realism.
\end{abstract}


\maketitle

\section{Introduction}\label{sec:intro}

Henry~R.\ Stapp~\cite{stapp77} once described the work of John~S.\
Bell~\cite{bell64} as ``the most profound discovery of science''.
Indeed, the work of Bell showed that our
intuition
that the world  should be local realistic is incorrect, thus changing
our perception of the physical world, perhaps to the same extent as
Isaac Newton's work on classical dynamics and Albert Einstein's
work on relativity. Albert Einstein, Boris Podolsky and Nathan Rosen
(EPR), defenders of the local realistic viewpoint, argued that
quantum mechanics is not a complete theory for it does not contain
every element of physical reality in its formalism~\cite{epr35}.
Bell showed that these exact same elements of reality, weaved into a
local model of Nature,
lead to a theory which contradicts
the predictions of quantum mechanics. The experimental verification
of Bell's predictions~\cite{clauser,aspect} gives us strong evidence
that Nature indeed does not have a local realistic description.

More recently, a new kind of refutation of the local realistic
viewpoint has arisen~\mbox{\cite{hr83,ghz89,mermin90a,hardy92}}.
These local-hidden-variable no-go theorems are also called ``Bell
theorems without inequalities''. Like standard Bell theorems, these
experiments must be repeated for many runs in order to rule out a
local realistic viewpoint (if we observe a single successful run, we
cannot conclude anything except maybe that quantum mechanics is
right or that a local-hidden-variable (LHV) model was lucky!) but
usually less runs are required in order to reach the same confidence
level as for standard Bell theorems. Another advantage is that the
proof that no LHV model can reproduce the quantum correlations is
usually much more elegant and simple. Instead of only showing that
no LHV model can reproduce the correlations predicted by quantum
mechanics (as is the case for standard Bell theorems), Bell theorems
without inequalities show that an LHV model which is to attempt a
simulation of quantum mechanics will run into a contradiction with
itself~\cite{methot05}. Most of these Bell theorems can be recast
into the framework of \emph{pseudo-telepathy}~\cite{bbt04a,bbt04b}.
In the pseudo-telepathy paradigm, proofs of non-locality are
presented in the form of games. These games consist of questions
given to space-like separated players who must give answers
satisfying a certain relation with the questions. We say that a game
which cannot be won with certainty by classical players (who share
common classical information), whereas it can be won with certainty
by quantum players (who share entanglement), is a pseudo-telepathy
game. In other words, any LHV model that is to attempt to simulate
the quantum correlations will, once in a while, output something
that is forbidden according to quantum mechanics. There exists a
Bell theorem without inequalities that cannot be transformed into
pseudo-telepathy: Lucien Hardy's theorem~\cite{hardy92}. Hardy's
argument uses a pair of non-maximally entangled qubits, and such a
state cannot produce correlations that yield a pseudo-telepathy
game~\cite{bmt04}.

In the last few months, new scenarios that can be cast into the
framework of two-player pseudo-telepathy games have been
proposed~\cite{ghz05a,ghz05b,cabelloa,cabellob}. The authors claim
that they present new proofs against local realism. Although the
equations that they present are mathematically correct, it is not
possible to interpret them in such a way as to rule out all LHV
models for the proposed experiments.
Our work aims to clarify this situation.  The
present paper is divided such that we first discuss candidates for
pseudo-telepathy games that use entanglement swapping in Section~II.
In Section~III, we analyse the treatment of LHVs which are time-like
separated. Before concluding, we finish with a discussion on
elements of reality in Bell experiments in Section~IV.

\section{Entanglement swapping}\label{sec:ghz}

Daniel~M.\ Greenberger, Michael~A.\ Horne and Anton
Zeilinger~\cite{ghz05a,ghz05b}, recently proposed schemes based on
entanglement swapping that fit in the framework of
pseudo-telepathy.
Here, we present a simple proof that shows that there is an LHV model for \emph{any} two-participant protocol
based on entanglement swapping.
Thus, without going into the details of the scheme, we show that  the experiment of
Greenberger, Horne and Zeilinger cannot rule out local realism. Afterwards, we
show that even if we consider the three-participant version of the Greenberger, Horne and Zeilinger protocol,
it still admits an LHV model.

Recall that the following are the four Bell
states:
\begin{align}
\psiminus= &\, \textstyle{\frac{1}{\sqrt{2}}}\ket{01}-\textstyle{\frac{1}{\sqrt{2}}}\ket{10}, \\
\psiplus= &\, \textstyle{\frac{1}{\sqrt{2}}}\ket{01}+\textstyle{\frac{1}{\sqrt{2}}}\ket{10},\\
\phiminus= &\, \textstyle{\frac{1}{\sqrt{2}}}\ket{00}-\textstyle{\frac{1}{\sqrt{2}}}\ket{11},\\
\phiplus= &\,
\textstyle{\frac{1}{\sqrt{2}}}\ket{00}+\textstyle{\frac{1}{\sqrt{2}}}\ket{11}.
\end{align}
In the entanglement swapping scheme, Bob shares a copy of the  state
$\psiminus$ with both Alice and Charlie, while Alice and Charlie are
not entangled. In order to \emph{swap} his entanglement, Bob then
measures his two qubits in the Bell basis. Before the measurement,
the state of the global system is, up to local unitaries,
\begin{equation}\label{eq:vwstate}
 \frac{1}{2}\psiminus_{\text{B}}\psiminus_{\text{AC}} +
\frac{1}{2}\psiplus_{\text{B}}\psiplus_{\text{AC}} + \\
 \frac{1}{2}\phiminus_{\text{B}}\phiminus_{\text{AC}} +
\frac{1}{2}\phiplus_{\text{B}}\phiplus_{\text{AC}},
\end{equation}
 where the first two qubits belong to Bob, the third to Alice
and the last to Charlie. After Bob's measurement, Alice and Charlie
are therefore left in a Bell state. The fact that the entanglement
between Alice and Charlie comes from particles that \emph{never}
interacted means that they do not share LHVs (but note that any experimental setup
that uses this hypothesis would have to be extremely well
orchestrated in order to ensure that Alice and Charlie were
\emph{never} in a situation where they could communicate). The
argument presented in~\cite{ghz05a,ghz05b} then goes on to analyse
the correlations of a Bell state as to whether they can be simulated
by an LHV model where Alice's and Charlie's particles \emph{do not}
share any variables. The assumption made here is that whatever
happens in Bob's lab is of no consequence. The reason given is that
the LHVs of the particles belonging to Alice and Charlie cannot depend on each other
and cannot depend on what happened in Bob's lab.
In the given
interpretation of the experimental scheme, Bob's lab can be thrown
into a black hole for all it matters.

Is this argument valid?
The answer is no. If Bob's knowledge of the outcome of the Bell
measurement is lost, Alice and Charlie are left with a mixture of
all the Bell states, each with equal probability. Obviously, this
state is the totally mixed state and it is \emph{not} entangled.
Therefore, Alice's and Charlie's answers will not be correlated in
any fashion. A simple LHV model can then simulate measurements on
Alice's and Charlie's particle: output at random! (while taking into
account that for a general POVM on a totally mixed state, not every
POVM element will be produced with equal probability, and adjusting
the marginal probabilities accordingly). Hence, without even
considering the specific measurements that are performed in the
experiment, we conclude that \emph{any} scheme with two participants
that is based on entanglement swapping admits an LHV model.

What if, instead of sending Bob into a black hole, we take into
consideration his outcome? If we know Bob's measurement outcome,
then we know the actual Bell state that is shared between  Alice and
Charlie.
We will now rewrite the experiment of~\cite{ghz05a} in the language
of quantum information, and consider the case where \emph{Bob's
measurement results are taken into consideration}. We show that this
experiment also admits an LHV model that simulates the correlations,
and that this is due to the fact that Bob shares LHVs with Alice
\emph{and} with Charlie.

Here is the scheme that we consider: Bob does a Bell state
measurement on the state described in Equation~\eqref{eq:vwstate}.
His outcome, say~$b$, is therefore one of the four Bell states.
Alice and Charlie are now left in a Bell state that depends on Bob's
measurement outcome. They are then both asked to perform the same
measurement: either in the standard basis (standard von Neumann
measurement, or $\sigma_z$) or in the  Hadamard basis (sending
$\ket{0}\rightarrow (\ket{0}+\ket{1})/\sqrt{2}$ and
$\ket{1}\rightarrow (\ket{0}-\ket{1})/\sqrt{2}$, followed by a von
Neumann measurement, or $\sigma_x$). Let $a \in\{-1,1\}$ be Alice's
outcome and~$c \in \{-1,1\}$ be Charlie's outcome. Alice and Charlie
each output a single bit ($-1$ or $1$), but to ease the notation we
will denote the outcomes $a_+$ and~$c_+$ if the measurements were
performed in the standard basis and $a_\times$ and~$c_\times$ if the
measurements were performed in the Hadamard basis. Depending on the
state that they share after Bob's measurement, their results will
either be correlated~($a\cdot c =1$) or anti-correlated~($a\cdot c
=-1$). Table~\ref{table:GHZgame} gives the measurements outcomes
that are predicted by quantum mechanics. Note also, that according
to these predictions, the local outcomes of Alice, Bob, and Charlie
are uniformly distributed.
\begin{table}[h]
\begin{center}
\begin{tabular}{|c|c|c|}
  \hline
  $b$ & $a_+ \cdot c_+$ &  $a_\times \cdot c_\times$  \\ \hline
  $\phiplus$ & $\phantom{-}1$ & $\phantom{-}1$  \\
  $\phiminus$ & $\phantom{-}1$ & $-1$  \\
  $\psiplus$ & $-1$ & $\phantom{-}1$  \\
  $\psiminus$ & $-1$ & $-1$  \\
  \hline
\end{tabular}
\caption{Measurement outcomes} \label{table:GHZgame}
\end{center}
\end{table}
Recall that we are in a scenario where Alice and Charlie do not
share hidden variables. At first sight it seems reasonable to think
that the correlations of Table~\ref{table:GHZgame} cannot be
fulfilled. However, Alice and Charlie \emph{are} allowed to share
variables with Bob. We will now show how they can exploit this to
reproduce the predictions of quantum mechanics using only LHVs.

Alice, Bob and Charlie share four LHVs that each take the value $-1$ or $1$ independently and with equal probability.
We denote these values by $\lambda_{a_+}, \lambda_{a_\times}, \lambda_{c_+}$ and  $\lambda_{c_\times}$. When
challenged to output the result of a measurement, Alice answers~$\lambda_{a_+}$ if the measurement is
in the standard basis and~$\lambda_{a_\times}$ otherwise. Charlie does the same, answering~$\lambda_{c_+}$
and~$\lambda_{c_\times}$ depending on his measurement. In order to give an answer that is consistent
 with table~\ref{table:GHZgame}, all that Bob needs to do is compute the values~$\lambda_{a_+} \cdot \lambda_{c_+}$
 and~$\lambda_{a_\times} \cdot \lambda_{c_\times}$. He then outputs the Bell state that he find in the corresponding
 row of Table~\ref{table:GHZgameLHV}. It is easy to see that this strategy that uses four bits of shared randomness
 satisfies all the conditions of Table~\ref{table:GHZgame} and that in addition, the local statistics correspond to those
 predicted by quantum mechanics. This technique works regardless of the order in which the participants are
 required to answer.
\begin{table}[h]
\begin{center}
\begin{tabular}{|c|c|c|}
  \hline
   $\lambda_{a_+} \cdot \lambda_{c_+} $ &  $\lambda_{a_\times} \cdot \lambda_{c_\times}$ & $b$  \\ \hline
     1 & $1$ & $\phiplus$  \\
    $1$ & $-1$ &  $\phiminus$  \\
    $-1$ & $1$ & $\psiplus$  \\
    $-1$ & $-1$ & $\psiminus$ \\
  \hline
\end{tabular}
\label{table:GHZgameLHV} \caption{LHV simulation}
\end{center}
\end{table}

The  technique that we have used is reminiscent to \emph{postselection}: according to the answers that
Alice and Charlie are to give, Bob selects an appropriate measurement outcome. This is
similar to~\cite{gg02},
which, surprisingly, rules out the results presented many years
later in~\cite{ghz05b}.
We can formulate a similar argument against the Bell
theorem presented by Ad\'an Cabello in~\cite{cabelloc}, as well as
the one presented by Zeng-Bing Chen, Yu-Ao Chen and Jian-Wei Pan~\cite{pan}.
In all these cases, postselection is
used in order to generate quantum correlations that cannot be
produced by any LHV model. These experiments omit to consider the
possibility that, as we have shown above, an LHV model can use postselection  to its
advantage.

\section{Inside the light cone}\label{sec:cabello}

What if we consider particles that were created in space-like
separated regions of space-time that are later brought together?
Could experiments performed on such particles and analysed with the
hypothesis that these particles cannot share any LHVs be convincing?
In~\cite{cabelloa,cabellob}, it is argued that different physical
quantities of a particle are elements of reality in the EPR sense.
Then, the values of these observables are analysed as being
independent since they \emph{are} elements of reality. One might be
tempted to think that these assumptions are reasonable, however they
are not. While creating the particles in space-like separated
regions will ensure that they do not share any LHVs at that point,
we cannot assume that this property is conserved for the entire
evolution of the system. In an LHV model, we do require that what
happens to a particle outside the light cone of another cannot have
any influence on the latter, but we can allow the particles to
constantly broadcast information that is secret to us (hidden
travelling information) which travels at the speed of light in all
directions. Therefore, once we bring a particle in the forward light
cone of the other, its LHVs can be influenced by those of the other
particle. This type of model is consistent with the local realistic
viewpoint and invalidates the assumption that the LHVs will stay
independent. This argument applies \emph{mutatis mutandis} to the
assumption that different observables, which are elements of
reality, do not share LHVs.

We now give a brief summary of the scheme proposed in~\cite{cabelloa}, which uses
the four-qubit state:
\begin{equation}\label{eq:psistate}
|\psi\rangle=
 \frac{1}{2}\big( 
  |0\rangle_{1}|0\rangle_{2}|0\rangle_{3}|0\rangle_{4}
 +|0\rangle_{1}|1\rangle_{2}|0\rangle_{3}|1\rangle_{4}+ 
 |1\rangle_{1}|0\rangle_{2}|1\rangle_{3}|0\rangle_{4}
 -|1\rangle_{1}|1\rangle_{2}|1\rangle_{3}|1\rangle_{4} \big).
\end{equation}
Qubits $1$ and $2$ belong to Alice and
qubits
 $3$ and $4$ to Bob.
Now consider the following three measurements $X_j, Y_j$ and $Z_j$,
performed individually on qubits $j$ ($j=1\ldots4$):
\begin{eqnarray}
X_{j} &= | 0 \rangle_{j} \langle 1 |
         + | 1 \rangle_{j} \langle 0 | \nonumber \\
Y_{j} &= i (| 1 \rangle_{j} \langle 0 |
         - | 0 \rangle_{j} \langle 1 |) \label{measure}\\
Z_{j} &= | 0 \rangle_{j} \langle 0 |
         - | 1 \rangle_{j} \langle 1|\,. \nonumber
\end{eqnarray}
Each of these measurements has two possible outcomes,
$+1$ and $-1$. Let the outcome of measurement $X_{j}$ be written
$x_{j} \in \{ +1,-1 \}$, and similarly for $Y_j$ and $Z_j$. Quantum
mechanics tells us that when appropriate measurements are made on
state  $|\psi \rangle$, the following four equalities
always hold:
\begin{align}
x_{1}  &=\phantom{-}x_{3}z_{4}, \label{1}\\
y_{1} &=-y_{3}z_{4}, \label{2}\\
x_{1}x_{2}&=\phantom{-}y_{3}y_{4}, \label{3} \quad \text{and}\\
y_{1}x_{2}&=\phantom{-}x_{3}y_{4}   \label{4}
\end{align}
In the scheme proposed in~\cite{cabelloa}, Alice is asked one of two
possible questions:
\begin{quote}
1a. What are $x_{1}$ and $x_{2}$? \\
2a. What are $y_{1}$ and $x_{2}$?
\end{quote}
Bob is independently asked one of four possible questions:
\begin{quote}
1b. What are $x_{3}$ and $y_{4}$? \\
2b. What are $x_{3}$ and $z_{4}$? \\
3b. What are $y_{3}$ and $y_{4}$? \\
4b. What are $y_{3}$ and $z_{4}$?
\end{quote}
The challenge that Alice and Bob face is to provide answers to these
questions such that Equations~\eqref{1}--\eqref{4} are satisfied.
Although it is shown in~\cite{cabelloa} that there is an element of
reality corresponding to each measurement result, it is also
possible that particles inside the same light cone can exchange
unlimited information. Therefore, measurements on separate particles
can be seen, for LHV model purposes, as one measurement on a global
system.

We now give an explicit LHV model that perfectly mimics the
predictions of quantum mechanics for the above scenario. Alice and
Bob share two random variables,  $\lambda_1$ and $\lambda_2$.
Regardless of the question she is asked, Alice always answers
``$\lambda_1$'' and ``$\lambda_2$''. Bob's strategy is to first flip
a fair coin. The outcome ($-1$ or $1$) of this coin flip is Bob's
first answer, call it $b_1$.  Bob then computes his second answer,
$b_2$ according to Table~\ref{table:CabellogameLHV} by using the
information that he has: the question that he was asked,
$\lambda_1$, $\lambda_2$ and $b_1$.
\begin{table}[h]
\begin{center}
\begin{tabular}{|c|c|}
  \hline
  question & $b_2$ \\ \hline
  1b & $\lambda_1 \cdot \lambda_2 \cdot b_1$ \\
  2b & $\lambda_1 \cdot b_1$  \\
  3b & $\lambda_1 \cdot \lambda_2 \cdot b_1$  \\
  4b &  $-\lambda_1 \cdot b_1$ \\
  \hline
\end{tabular}
\label{table:CabellogameLHV} \caption{Bob's strategy in the LHV
model for Cabello's game}
\end{center}
\end{table}
It is interesting to point out that our LHV model not only satisfies
the rules of Equations~\eqref{1}--\eqref{4}, but also reproduces the
predictions of quantum mechanics perfectly: it is easy to see that
in the LHV model, the local outcomes of Alice and Bob are uniformly
distributed, and that this corresponds exactly to the predictions of
quantum mechanics!

A similar argument can be used to
show an LHV model to simulate the experiment in~\cite{cabellob}. At
this point, it is important to stress that this LHV model is not a
contextual model in the usual sense of the term. Non-contextuality
has to do with the choice of output in a given POVM~\cite{ks67},
while here the ``context'' is which POVM is done on what particle.
This model is not contextual but uses hidden traveling information
between the particles and is of course consistent with a local
realistic viewpoint.

The argument presented by Cabello does rule out a certain class of
LHV models, those that do not use hidden traveling information, also
called the EPRLER model by Cabello. However, it does not rule out
every LHV model. There is of course a simple solution to make the
equations given in~\cite{cabelloa,cabellob} physically meaningful.
We keep the elements of reality in space-like separated regions of
space and give them to new players. We can thus convert the game
presented in~\cite{cabelloa,cabellob} into  convincing experimental
proposals~\cite{bm06}.

\section{Discussion and conclusion}

Since Bell's 1964 discovery, new Bell experiments  have continuously
been proposed. The goal of such experiments is to demonstrate
experimentally the nonlocality of the world in which we live. In
order to circumvent imperfections in the laboratory setting, new
experiments are proposed to close experimental loopholes, one of the
most notorious being the \emph{detection loophole}~\cite{dlh}. But
as we have demonstrated, not all Bell experiments are created
equally, and a careful analysis is required in order to verify the
validity of the proposed experiments. The  papers that we have
analysed here have something in common: they start by arguing for
 the existence of elements of reality and then base
their analysis of the experiment on these elements of reality.
However, the existence or independence of these elements of reality
is not tested in the final experimental setup. We believe that this
is what sets these experiments apart from others and that allows an
LHV model that explains the experiment.

One must also be careful using arguments that concern elements of
reality. Einstein, Podolsky and Rosen gave a criterion to recognize
elements of reality~\cite{epr35}:
\begin{quotation}
\emph{If, without in any way disturbing a system, we can predict with
certainty (i.e., with probability equal to unity) the value of a
physical quantity, then there exists an element of physical
reality corresponding to this physical quantity.}
\end{quotation}
\noindent It cannot be stressed enough that this criterion is
``regarded not as a necessary, but merely as a sufficient, condition
of reality''\cite{epr35}. Said differently, not every element of
reality can necessarily be measured without disturbing the system.
Otherwise, EPR would have claimed, after showing that momentum and
position can have simultaneous reality, that the Heisenberg
uncertainty relation can be violated~\cite{methot06}!

In order to propose meaningful experiments, it is
useful to use a higher level of abstraction  to analyse the
scenario: by placing the proposed experiments in the framework of
pseudo-telepathy, we have been able to show that an LHV model can
explain the results of the experiments. In fact, we believe that there is
much to gain by studying nonlocality in an adversarial context:
when analysing nonlocality proofs,
one should be just as paranoid about
Nature cheating our senses as are cryptographers about the security of a
protocol against attacks from a malicious adversary.

\section*{Acknowledgements}
We would like to thank Ad\'an Cabello, Hilary
Carteret, Nicolas Gisin, Alain Tapp and Jonathan Walgate for very
stimulating discussions and S\'ebastien Gambs for helpful comments
on the manuscript. This work is supported in part by Canada's {\sc
Nserc}.


\begin{thebibliography}{99}

\bibitem{stapp77}
H.\,R.~{Stapp}, ``Are superluminal connections necessary?''
\textit{Nuovo Cimento B} \textbf{40}, 191--204, 1977.

\bibitem{bell64}
J.\,S.~{Bell}, ``On the {E}instein-{P}odolsky-{R}osen paradox'',
\textit{Physics} \textbf{1}, 195--200, 1964.

\bibitem{epr35}
A.~{Einstein}, B.~{Podolsky} and N.~{Rosen}, ``Can
quantum-mechanical description of physical reality be considered
complete?'' \textit{Physical Review Letters} \textbf{47}, 777--780,
1935.

\bibitem{clauser}
J.\,S.~{Freedman} and J.\,F.~{Clauser}, ``Experimental Test of Local
Hidden-Variable Theories", \textit{Physical Review Letters}
\textbf{28}, 938--941, 1972.

\bibitem{aspect}
A.~{Aspect}, P.~{Grangier} and G. {Roger}, ``Experimental tests of
realistic local theories via {B}ell's theorem'', \textit{Physical
Review Letters} \textbf{47}, 460--463, 1981.

\bibitem{hr83}
P.~{Heywood} and M.~L.~G.~{Redhead}, ``Nonlocality and the
{K}ochen-{S}pecker paradox'', \textit{Foundations of Physics}
\textbf{13}, 481--499, 1983.

\bibitem{ghz89}
D.\,M.~{Greenberger}, M.~A.~{Horne} and A.~{Zeilinger}, in {\it
Bell's Theorem, Quantum Theory, and Conceptions of the Universe},
edited by M. Kafatos (Kluwer Academic, Dordrecht), pages~69--72,
1989.

\bibitem{mermin90a}
N.\,D.~{Mermin},  ``Quantum mysteries revisited'', \textit{American
Journal of Physics} \textbf{58}: 731--743, 1990.

\bibitem{hardy92}
L.~{Hardy},  ``Quantum mechanics, local realistic theories, and
{L}orentz-invariant realistic theories'', \textit{Physical Review
Letters} \textbf{68}: 2981--2984, 1992.

\bibitem{methot05}
A.\,A.~{M{\'e}thot},``On local-hidden-variable no-go theorems'',
\textit{Canadian Journal of Physics}, to appear. Available at
\url{http://arxiv.org/quant-ph/0507149}, 2005.

\bibitem{bbt04a}
G.~{Brassard}, A.~{Broadbent} and A.~{Tapp}, ``Quantum
pseudo-telepathy'', \textit{Foundations of Physics} \textbf{35},
1877--1907, 2005.

\bibitem{bbt04b}
G.~{Brassard}, A.~{Broadbent} and A.~{Tapp}, ``Recasting {M}ermin's
multi-player game into the framework of pseudo-telepathy'',
\textit{Quantum Information and Computation} \textbf{5}: 538--550,
2005

\bibitem{bmt04}
G.~{Brassard}, A.\,A.~{M\'ethot} and A.~{Tapp}, ``Minimal bipartite
state dimension required for pseudo-telepathy'', \textit{Quantum
Information and Computation} \textbf{5}: 275--284, 2005.

\bibitem{ghz05a}
D.\,M.~{Greenberger}, M.~{Horne} and A.~{Zeilinger}, ``A {B}ell
theorem without inequalities for two particles, using efficient
detectors'', available at \url{http://arxiv.org/quant-ph/0510201},
2005.

\bibitem{ghz05b}
D.\,M.~{Greenberger}, M.~{Horne} and A.~{Zeilinger}, ``A {B}ell
theorem without inequalities for two particles, using inefficient
detectors'', available at \url{http://arxiv.org/quant-ph/0510207},
2005.

\bibitem{cabelloa}
A.~Cabello,``Stronger two-observer all-versus-nothing violation of
local realism'', \textit{Physical Review Letters} \textbf{95}:
210401,  2005.

\bibitem{cabellob}
A.~Cabello,   ``Loophole-free {B}ell's experiment based on
two-photon all-versus-nothing violation of local realism'',
\textit{Physical Review A}, \textbf{72}: 050101(R), 2005.

\bibitem{bm06}
A.~{Broadbent}, H.~{Carteret}, A.\,A.~{M\'ethot} and J.~{Walgate},
``On the logical structure of Bell theorems without inequalities'',
available at \url{http://arxiv.org/quant-ph/0512201}, 2005.

\bibitem{pan}
Z.-B.~Chen, Y.-A.~Chen, J.-W.~Pan, ``All-versus-nothing violation of
local realism by swapping entanglement'', available at
\url{http://arxiv.org/quant-ph/0505178}.

\bibitem{gg02}
N.~Gisin and B.~Gisin, ``A local variable model for entanglement
swapping exploiting the detection loophole'', \textit{Physics
Letters A}, \textbf{297}: 279--284, 2002.

\bibitem{cabelloc}
A.~Cabello, ``Violating Bell's Inequality Beyond Cirel'son's
Bound'', \textit{Physical Review Letters}, \textbf{88}: 060403,
2002.


\bibitem{ks67}
S.~{Kochen} et E.~{Specker}, ``The problem of hidden variables in
quantum mechanics'', \textit{Journal of Mathematical Mechanics}
\textbf{17}: 59--87, 1967.

\bibitem{dlh}
P.~Pearle, ``Hidden-Variable Example Based upon Data Rejection'',
\textit{Physical Review D}, \textbf{2}: 1418–-1425, 1970.

\bibitem{methot06}
A.\,A.~{M\'ethot}, ``Can quantum-mechanical description of physical
reality be considered
 \textit{correct}'', in preparation.


\end{thebibliography}
\end{document}